\documentclass[a4paper,
               keeplastbox,   
               ]{jacow}
\usepackage{pdfpages,multirow,ragged2e} %
\makeatletter%
	\ifboolexpr{bool{xetex}}
	 {\renewcommand{\Gin@extensions}{.pdf,%
	                    .png,.jpg,.bmp,.pict,.tif,.psd,.mac,.sga,.tga,.gif,%
	                    .eps,.ps,%
	                    }}{}
\makeatother

\ifboolexpr{bool{xetex} or bool{luatex}} 
 {}                                      
 {\usepackage[utf8]{inputenc}}           

\usepackage[USenglish]{babel}
\usepackage{tablefootnote}

\ifboolexpr{bool{jacowbiblatex}}%
 {%
  \addbibresource{jacow-test.bib}
  \addbibresource{biblatex-examples.bib}
 }{}
\begin{document}

\title{Coupled bunch stability with variable filling patterns in PETRA~IV}

\author{S.~A.~Antipov\thanks{Sergey.Antipov@desy.de}, C. Li\\ Deutsches Elektronen-Synchrotron DESY, Notkestr. 85, 22607 Hamburg, Germany}
	
\maketitle

\begin{abstract}
   The PETRA IV upgrade project is aiming at building a 6~GeV diffraction-limited light source.  The storage ring’s off-axis accumulation injection scheme will allow generating a wide range of filling patterns for the needs of photon science users. To preserve high beam quality and low transverse emittances it is imperative to ensure beam stability against collective effects. In this paper we investigate the impact of different filling patterns on the coupled-bunch stability in the ring using a semi-analytical Vlasov solver.
\end{abstract}

\section{Introduction}

The PETRA~IV~\cite{bib:CDR} envisions a state-of-the art storage ring that will provide electron beams of unmatched brightness and quality for photon science experiments. Achieving and maintaining high brightness beams in operation requires careful control over potential collective instabilities that might arise due to self-induced wakefields. While in practice the instabilities can be mitigated by several mechanisms, such as chromaticity, feedbacks, and synchrotron radiation damping, the most sound approach is to ensure a significant safety margin during the design stage and reduce the machine impedance as much as possible. 

The large dynamic aperture of PETRA~IV hybrid H6BA lattice together with a long Touscheck lifetime allow operating the ring in an off-axis injection and accumulation mode. This is achieved by a set of fast injection stripline kickers~\cite{gregor} capable of injecting a single bunch in any of the 3840 2~ns RF buckets without distorting the stored beam. The fast single bunch off-axis injection allows creating and accumulating a large variety of filling pattern in the storage ring. 

As bunch separation and bunch charge affect significantly the strength of long-range wakes felt by the bunches in the beam, it is important to verify that all the filling schemes, both baseline and potential, could be operated with a sufficient safety margin.

\section{Filling patterns}
PETRA~IV is capable of creating a wide range of filling patterns. The two baseline filling schemes are a 200~mA Brightness mode with 1920 bunches and a 80~mA Timing mode with 80 bunches. Both schemes utilize uniform bunch spacing. Other, non-baseline patterns are also on the table, such as a 2~ns Brightness mode, a 1600 bunch Brightness mode from the CDR, a Hybrid pattern\footnote{Here we limit ourselves to filling patterns with bunches of equal charge. In reality a Hybrid pattern may use high charge `guarding' bunches at the ends of the train. Simulating the guarding bunches is beyond the capabilities of our computer code.} with only $7/8$ of the ring filled, or a 40-bunch Timing mode. Table~\ref{tab:FS} summarises the main features of different bunch patterns.

\begin{table}[h]
   \centering
   \caption{Parameters of the PETRA~IV lattice (IDs closed).}
   \begin{tabular}{lr}
       \toprule
       \textbf{Parameter} & \textbf{Value}\\
       \midrule
           Electron energy & 6~GeV\\
           Revolution freq. & 130.2~kHz\\
           Tunes: x,y,z & 135.22, 86.20, 0.005\\
           Momentum comp. & $3.3\times 10^{-5}$\\
           Synch. damp. times: x,y,z & 18, 22, 13~ms\\                    
        \bottomrule
   \end{tabular}
   \label{tab:beam_par}
\end{table}

\begin{table}[h]
   \centering
   \caption{PETRA~IV baseline Brightness mode and some alternative scenarios.}
   \begin{tabular}{lrrrr}
       \toprule
       \textbf{Filling} & \textbf{No. bun.} & \textbf{Spacing} & \textbf{Sep.} & \textbf{Current} \\
       \midrule
           Baseline & 1920 & Uniform & 4~ns & 200~mA\\
           CDR & 1600 & 20b+4e & 4~ns & 200~mA\\
           Hybrid & 1680 & 7/8 & 4~ns & 200~mA\\
           2~ns & 3840 & Uniform & 2~ns & 200~mA\\    
        \bottomrule
   \end{tabular}
   \label{tab:FS}
\end{table}

\section{\label{sec:NHT}Nested Head-Tail Formalism}

Transverse modes of a coherent motion can described by a linearized Vlasov equation, assuming no significant diffusion or coupling effects. 
In the polar coordinates $(y, p_y)\rightarrow(q, \theta)$, $(z, \delta)\rightarrow(r, \phi)$ it reads:
\begin{equation} \label{eq:2}
	\frac{\partial\psi}{\partial s} + \frac{\omega_\beta}{c}\frac{\partial\psi}{\partial\theta} + \frac{F_y(z,s)}{E}\frac{\partial\psi}{\partial p_y} + \frac{\omega_s}{c}\frac{\partial\psi}{\partial\phi} = 0,
\end{equation}
where $\omega_{\beta,s}$ are the betatron and synchrotron frequencies, $E$ is the beam energy, and $F_y$ is the transverse wake force generated by the dipole moment of the beam, and $c$ is the speed of light. Here and further we consider a relativistic beam, i.e. its Lorentz factor $\gamma \gg 1$. Eq.~(\ref{eq:2}) can be solved perturbatively, assuming a small perturbation on top of the stationary distribution, and discretizing the distribution function on a set of radial rings \cite{bib:NHT}. Further assuming that the long-range wake is flat, i.e. its variation over the bunch length can be neglected~\cite{bib:Burov_Transverse_Modes}, one arrives at a simple set of linear equations eigenmode vectors $X$ and their complex frequencies $\Omega$ for a given coupled-bunch mode number $\mu$:
\begin{equation} \label{eq:CB}
	\begin{split}
		& \frac{\Delta\omega}{\omega_s} X = S X - iZ X - igF X + W^\mu F X,
  	\end{split}
\end{equation}
where $\Delta\omega = \Omega - \omega_\beta$ is the mode's complex frequency shift and $S$, $Z$, $F$ are synchrotron, single-bunch, and damper and $W^\mu$ represents the effect of the inter-bunch wake $W$ on the $\mu$-th coupled-bunch mode. $S$ denotes the matrix of harmonic oscillations inside the RF potential well:
\begin{equation} \label{eq:S}
S = l\delta_{lm}\delta_{\alpha\beta}.
\end{equation}
The transverse impedance $Z_1^{\perp}$ defines the single-turn impedance of the problem:
\begin{equation} \label{eq:Z}
Z = i^{l-m}\frac{\kappa}{n_r}\int_{-\infty}^{+\infty}Z_1^\perp(\omega^{\prime}) J_l(\chi^{\prime}_\alpha) J_{m}(\chi^{\prime}_\beta) d\omega^{\prime},
\end{equation}
where $\kappa = N r_0 c / 2\gamma\omega_\beta\omega_s T_0^2$ is the normalized intensity parameter, $n_r$ is the number of radial rings, and $\chi = \omega_\xi \tau$ is the head-tail phase advance for a given radius. $N$ is the number of particles per bunch, $r_0$ their classical radius, $T_0$ the revolution period, and $J_l$ are the $l$-th order Bessel functions of the first kind.

Finally, the damper term represents the response of a flat damper,  i.e. a damper that acts only on the center of mass of the beam, and whose kick is flat in the time domain. $g$ stands for the normalized damper gain in the units of $\omega_s$. Its matrix $F$ can be quickly derived from Eq.~(\ref{eq:Z}) after a substitution $Z(\omega) \approx \delta(\omega)$:
\begin{equation} \label{eq:F}
F = \frac{i^{m-l}}{n_r} J_l(\chi_\alpha) J_{m}(\chi_\beta).
\end{equation}
\subsection{Equidistant bunches}
Normally, one considers $M$ equidistant bunches. Then the coupled bunch term $W^\mu$ can be found from the rotational symmetry of the problem:
\begin{equation} \label{eq:W}
	W^\mu = 2\pi\kappa \sum_{k=1}^{\infty} W(-k s_0)\exp(2\pi i \nu_{k\mu}),
\end{equation}
where the wake $W$ is sampled at the bunch centers with the spacing of $s_0$, $\mu = 0,...M-1$, and $\nu_{k\mu} = k(\mu+\nu)/M$.
\subsection{Arbitrary bunch spacing}
If the bunches are not equidistant, one has to, first, solve for the proper coupled-bunch modes $w^\mu$
\begin{equation} \label{eq:W_arb}
	w^\mu Y^\mu = (W - igG)Y^\mu,
\end{equation}
where $W$ is a matrix of intra-bunch wakes 
\begin{equation} \label{eq:W_matrix}
	W_{ij} = 2\pi\kappa \sum_{k}W(-k C + \Delta s_{ji})
        \exp[2\pi i \nu (k + \Delta s_{ji} / C)]
  \end{equation}
and the summation goes from $k = 1$ to $\infty$ if the distance between $i$-th and $j$-th bunches $\Delta s_{ji} \geq 0$ and from $k = 0$ otherwise. $G$ is the damper matrix describing kick of the feedback system on the target bunch and its neighbours. For an ideal damper, which has no leakage of kick to neighboring bunches, $G = I$.

Thus the multi-bunch problem reduces to a set of single-bunch problems for each coupled-bunch mode $w^\mu$, which can be solved separately.
The Nested Head Tail (NHT) Vlasov solver \cite{bib:NHT, Antipov:2018aaw} has been modified to implement this procedure. The modified code allows performing rapid multiparametric scans over arbitrary filling patterns.

For practical filling patters the system Eq.~(\ref{eq:W_arb}) can be solved very efficiently since the bunch patters are usually relatively regular, hence many coefficients in $W$ are the
same and need to be computed only once. Furthermore, since wakes decay with distance, the resulting matrix is sparse, allowing finding the most prominent eigenvalues with iterative Arnoldi solvers\footnote{Another way to utilise the sparseness of the matrix is to use the lower rank approximation; see \cite{Lower_Rank} for details.}. All this allows efficiently computing the most unstable coupled-bunch modes even for systems as large as FCC with over $10^4$ bunches on a personal laptop.

\section{Impedance model}
\begin{figure}[!ht]
   \centering
   \includegraphics*[width=.95\columnwidth]{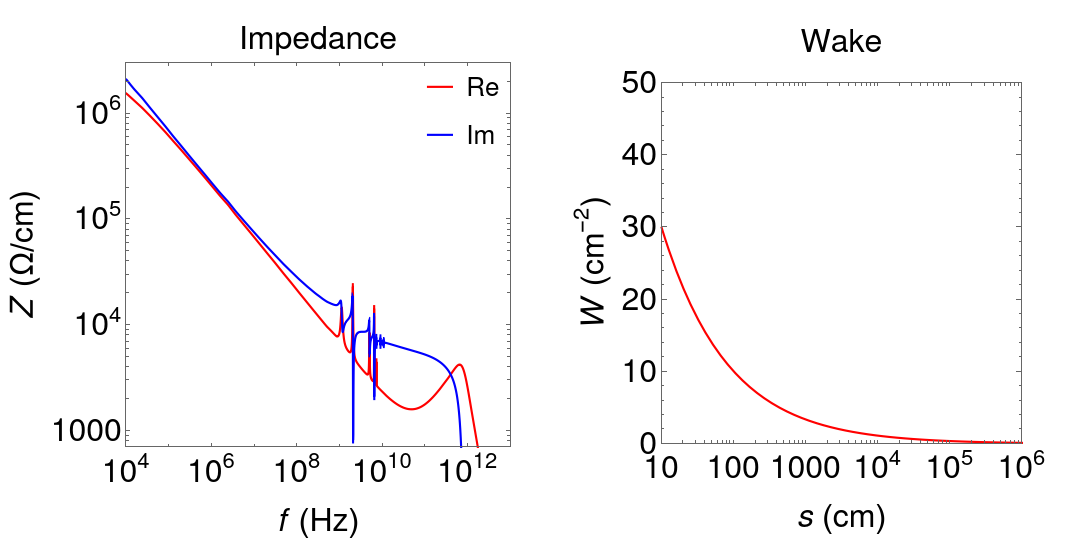}
   \caption{PETRA~IV impedance and long-range wake used for the studies.}
   \label{fig:Imp-Wake}
\end{figure}

The impedance model of the ring consists of the resistive wall contributions of its standard and ID vacuum chambers and multiple sources of geometric impedance: RF cavities, injection and feedback kickers, BPMs, current monitors, tapers, bellows, and radiation absorbers. The dominant resistive wall contribution of ring's vacuum chambers is modelled using the IW2D code~{\cite{bib:IW2D}}. All vacuum chambers are assumed to be coated with NEG (resistivity of 2~$\mu\Omega$m, according to RF measurements of coated tube samples~\cite{bib:NEG}).  As for the geometric impedance, at the time of writing of this paper many hardware components, such as striplines and tapers, are still in early design stages and are included in the model as effective broadband impedance. In practice, one can expect the total geometric impedance to be about or smaller than 30\% of the overall impedance budget~\cite{Chae:2018vqx}. Figure~\ref{fig:Imp-Wake} depicts the transverse impedance and wake used for the studies.

\begin{figure}[!ht]
   \centering
   \includegraphics*[width=.95\columnwidth]{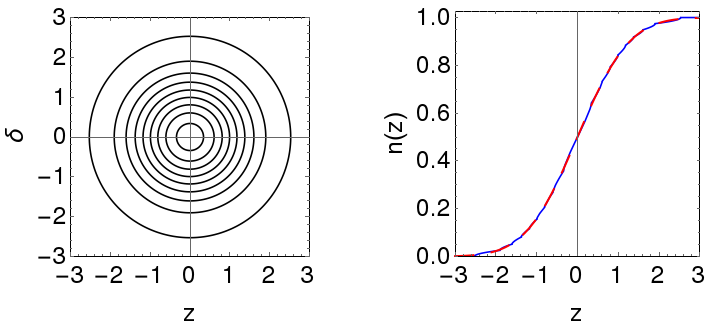}
   \caption{Longitudinal phase space distribution of the radial rings (left) and the corresponding normalized integrated charge density (right): the blue line depicts the integrated density, the dashed red - for a true Gaussian beam.}
   \label{fig:AB}
\end{figure}

In order to correctly treat the initial eigenvalue problem the number of rings has to reflect the wake properties. If the wake includes prominent high-frequency components the number of rings has to be increased such that the characteristic phase advance between two neighbors is small:
\begin{equation} \label{eq:crit}
	2\pi f\Delta \tau \ll 1,
\end{equation}
where $f$ is the characteristic frequency of the wake, and $\Delta\tau$ is the distance between the rings. For the PETRA wake, dominated by the resistive wall effects, a choice of $n_r = 9$ works reasonably well. Figure~\ref{fig:AB} shows the radial discretization.

\section{Numerical simulation}
\begin{figure}[!ht]
   \centering
   \includegraphics*[width=.7\columnwidth]{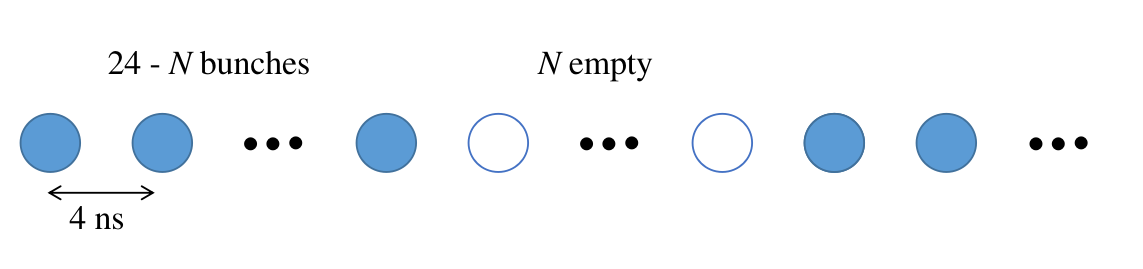}
   \caption{Setup of the filling scheme scan.}
   \label{fig:gap_setup}
\end{figure}

To assess the effect of gaps in filling patterns on beam dynamics we have conducted a series of parameter scans, varying the lengths of the gaps while keeping the bunch charge constant. The gaps varied in length from 0 (no gap) to 20 bunches every 24 bunches with 4~ns spacing~(Fig.~\ref{fig:gap_setup}). The simulation included effects of transverse feedback and chromaticity. The numerical setup included 21 azimuthal, 9 radial head-tail modes, and up to 1920 coupled-bunch modes. 

With the transverse feedback off there is a sizeable effect of gaps in the filling patterns on the instability growth rate. The difference is the most prominent at zero chromaticity and decreases with chromaticity. A sufficiently strong feedback was found to remove the dependence on the filling pattern. The instability growth rate depends only on the bunch charge, and not on the total beam current for all chromaticities except $\xi = 0$ with a 50-turn feedback. For comparison, the maximum damping rate of the PETRA~IV feedback will be 1/40 turns with an option to increase it further, if needed.

\begin{figure}[!ht]
   \centering
   \includegraphics*[width=.99\columnwidth]{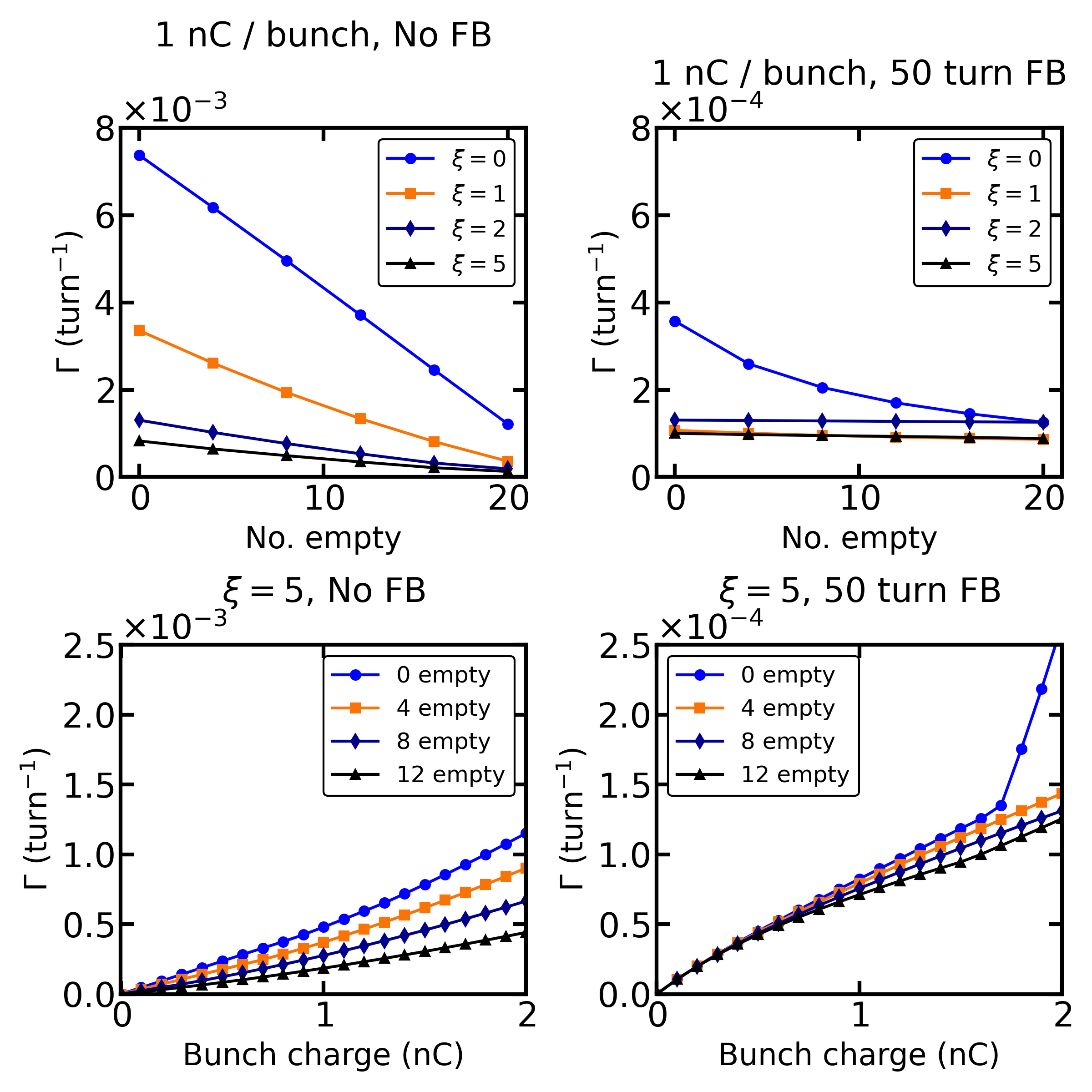}
   \caption{There is no significant difference in the coupled-bunch growth rate for different filling schemes with a sufficiently strong feedback and chromaticity. Growth rate of the most unstable mode as a function of number of empty buckets or bunch charge for different chromaticity and damper settings. The feedback gain is set to 0 in the left data and to 1/50 turns in the right data.}
   \label{fig:resutls}
\end{figure}

\section{Conclusion}

Irregularities, such as gaps, in filling patters can have an impact on transverse coupled-bunch dynamics in a storage ring. For the PETRA~IV machine, whose impedance is dominated by resistive wall effects, the details for the chosen filling pattern might vary the dynamics dramatically in the absence of couple-bunch feedback and at 0 chromaticity. At the same time, sufficiently strong feedback and chromaticity break coupled-bunch interaction, reducing the stability problem to, effectively, a single bunch case. Then the choice of the filling pattern has no significant impact on the transverse dynamics, rather it is mostly governed by the charge per bunch.


\ifboolexpr{bool{jacowbiblatex}}%
	{\printbibliography}%
	{%
	
	
}
\end{document}